\documentclass[a4paper]{jpconf}
\usepackage{graphicx}
\usepackage{amsfonts}
\usepackage{amsmath}
\usepackage{amssymb}
\begin{document}
\title{$f(R)$ brane cosmology}

\author{Mariam Bouhmadi-L\'opez}

\address{Centro Multidisciplinar de Astrof\'{\i}sica - CENTRA, Departamento de F\'{\i}sica, Instituto Superior T\'ecnico, Av. Rovisco Pais 1,
1049-001 Lisboa, Portugal}

\ead{mariam.bouhmadi@ist.utl.pt}

\begin{abstract}
Despite the nice features of the Dvali, Gabadadze and Porrati (DGP) model to explain the late-time acceleration of the universe, it suffers from some theoretical problems like the ghost issue. We present a way to self-accelerate the normal DGP branch, which is known to be free of the ghost problem, by means of an f(R) term on the brane action. We obtain the de  Sitter self-accelerating solutions of the model and study their stability under homogeneous perturbations.
\end{abstract}

\section{Introduction}

One of the most puzzling problems nowadays in physics  is the issue of the late-time acceleration of the universe \cite{Durrer:2008in}. A possible approach to tackle this problem is within the frame-work of self-accelerating universes \cite{dgp,Nojiri:2006ri,Capozziello:2007ec,BouhmadiLopez:2009db}; i.e. could  a modification of gravity at late-time and on large scale be the cause of the current inflationary phase of the universe? In other words, could this modification of gravity on large scales provide an \textit{effective negative pressure} that would fuel the late-time acceleration of the universe?

A possible approach to tackle these questions is the Dvali, Gabadadze and Porrati (DGP) scenario \cite{dgp}, which corresponds to a  five-dimensional (5D) model. In this model, our universe is a brane; i.e. a 4D hyper-surface, embedded in a flat space-time.  The DGP model has two types of solutions: the self-accelerating branch and the normal one. The self-accelerating brane is asymptotically de Sitter. This feature takes place without invoking any unknown dark energy component. On the other hand, the normal branch requires a dark energy component to accommodate the current observations \cite{Durrer:2008in}. Despite the nice features of the self-accelerating DGP branch, it suffers from serious  theoretical problems like the ghost issue \cite{Koyama:2007za}. In this paper we  propose a mechanism to self-accelerate the normal branch which is known to be free from the ghost issue \cite{Koyama:2007za}. This mechanism will be based on a modified Hilbert-Einstein action on the brane \cite{BouhmadiLopez:2009db} and the simplest gravitational option is an $f(R)$ term \cite{Capozziello:2007ec}.

\section{The model}

We consider a  brane, described by a 4D hyper-surface ($h$, metric g), embedded in a 5D bulk space-time ($\mathcal{B}$, metric 
$g^{(5)}$), whose action is given by 
\begin{eqnarray}
\mathcal{S} = \,\,\, \int_{\mathcal{B}} d^5X\, \sqrt{-g^{(5)}}\;
\left\{\frac{1}{2\kappa_5^2}R[g^{(5)}]\;\right\}
 + \int_{h} d^4X\, \sqrt{-g}\; \left\{\frac{1}{\kappa_5^2} K\;+\alpha f(R) + \mathcal{L}_m \right\}\,, \label{action}
\end{eqnarray}
where $\kappa_5^2$ is the 5D gravitational constant,
$R[g^{(5)}]$ is the scalar curvature in the bulk and $K$ the extrinsic curvature of the brane in the higher dimensional
bulk. On the other hand,  $R$ is the scalar curvature of the induced metric on the brane, $g$, and  $\alpha$ is a constant that measures the strength of the generalised induced gravity term $f(R)$ and has mass square units. Notice that therefore  the function $f(R)$ has mass square units.
 Finally, $\mathcal{L}_m$ corresponds to the matter Lagrangian of the brane. We will assume that the brane splits the bulk in two symmetric pieces. The previous action, includes as a particular case the DGP model \cite{dgp} for $f(R)=R$ and $\alpha=1/2\kappa_4^2$ where $\kappa_4^2$ is proportional to the 4D gravitational constant. It can be shown that the total energy density of the brane is conserved (we refer the reader to \cite{BouhmadiLopez:2009db} for more details). In particular, the energy density of matter on the brane is conserved.

In what follows, we consider a  homogeneous and isotropic brane. The matter sector on the brane can be described by a perfect fluid with energy density $\rho^{(m)}$ and pressure $p^{(m)}$,  where $\rho^{(m)}$ is conserved as we have  pointed above. On the other hand, an effective energy density and an effective pressure associated to the energy momentum tensor coming from the $f(R)$ term on action can be defined as follows \cite{BouhmadiLopez:2009db}
\begin{eqnarray}
\rho^{(f)}\label{rhof} &=&-2\alpha\left[3\left(H^2+\frac{k}{a^2}\right)f'-\frac12(Rf'-f)+3H\dot R f'' \right], \label{rhof}\\
p^{(f)}&=& 2\alpha\left\{\left(2\dot H+3H^2+\frac{k}{a^2}\right)f'-\frac12(Rf'-f) \left[\ddot R f''+(\dot R)^2f'''+2H\dot R f''\right]
\right\},\label{pf}
\end{eqnarray}
Notice that the definition of $\rho^{(f)}$ and $p^{(f)}$ is different from the standard 4D definition in $f(R)$ models \cite{BouhmadiLopez:2009db}. On the other hand, the energy density is conserved on the brane.

The modified Friedmann equation on the brane can be written as 
\begin{eqnarray}
3H^2= \frac{\kappa_5^4}{12}\rho^2.\label{friedmann}
\end{eqnarray}
While, the spatial component of Einstein equation can be expressed as       
\begin{eqnarray}
2\dot H+ 3H^2+\frac{k}{a^2}= -\frac{\kappa_5^4}{12}\rho(\rho+2p),
\label{ray}
\end{eqnarray}
where the  energy density $\rho$ and the pressure $p$ are defined as  
\begin{eqnarray}
\rho=\rho^{(m)}+\rho^{(f)}, \quad
p=p^{(m)}+p^{(f)}.
\label{rhop}
\end{eqnarray}

For simplicity, on equations (\ref{friedmann}) and (\ref{rhop}) we have used  the spatially flat chart of the brane.

\section{Self-accelerating branes}

A de Sitter universe is  the simplest cosmological solution that exhibits acceleration and therefore it is worthwhile to prove the existence of this solution in our model and study its stability. This would be a first step towards describing in a realistic way the late-time acceleration of the universe in an $f(R)$ brane-world model. This approach will also enable us to look for self-accelerating solutions on the modified normal DGP branch. So, in this section, we first obtain the fixed points of the model corresponding to a de Sitter space-time and then we study their stability under homogeneous perturbations. 

\subsection{Background solutions}

\begin{figure}[h]
\begin{center}
\includegraphics[width=7cm]{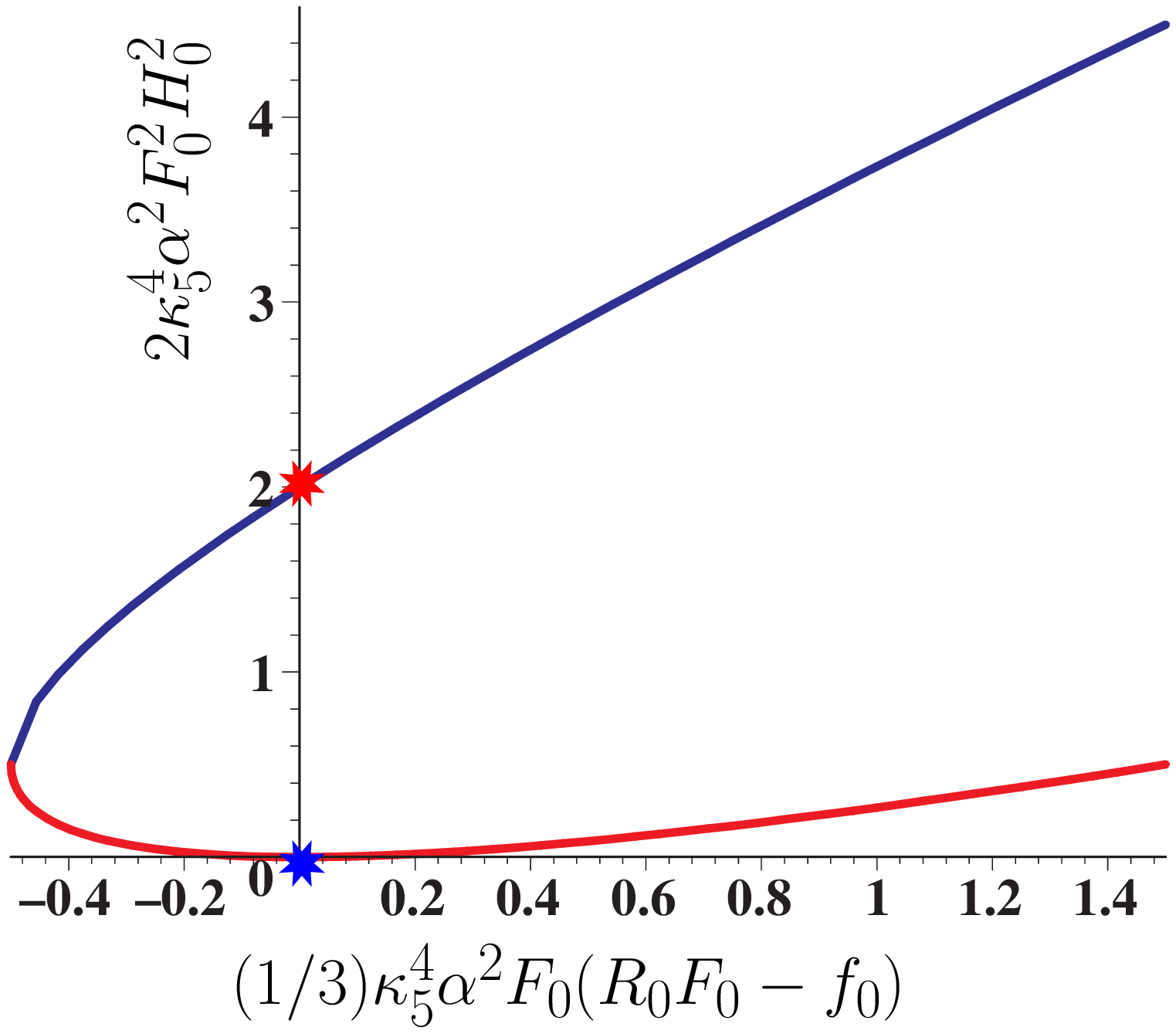}\hspace*{1.5cm}\includegraphics[width=7cm]{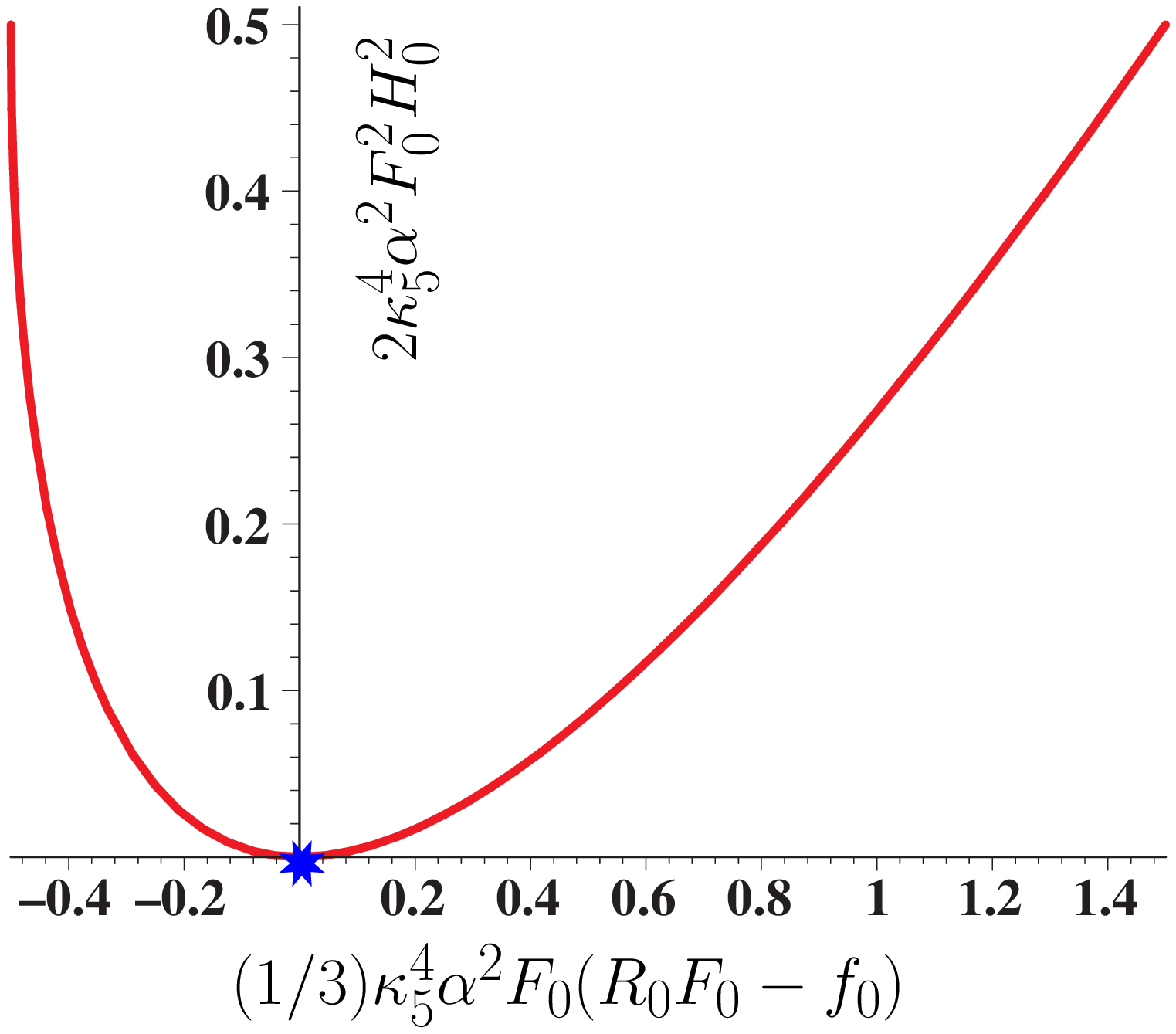}
\end{center}
\caption{The figure on the left shows the behaviour of the rescaled squared Hubble rate $2\kappa_5^4\alpha^2F_0^2H_0^2$ for the two branches that generalise the DGP solution versus the rescaled energy density $\rho^{(c)}$ defined as $\frac13\kappa_5^4\alpha^2F_0(R_0F_0-f_0)$. The blue star corresponds to the normal DGP branch which is flat. The red star corresponds to the self-accelerating DGP branch. On the other hand, the blue curve corresponds to the generalised (by the inclusion of the $f(R)$ term) self-accelerating branch, while the red curve corresponds to the generalised (by the inclusion of the $f(R)$ term) normal branch. The figure on the right corresponds to a zoom of the normal branch as it appears on the figure of the left.}
\label{ploth}
\end{figure}

In our model, the Hubble parameter  for de Sitter solutions can be expressed as\footnote{For a maximally symmetric brane in our model, the matter content of the brane behaves like a cosmological constant. As such a term can always be reabsorbed in the $f(R)$ term  we will disregard the matter content in our analysis of de Sitter branes.} 

\begin{eqnarray}\label{desitterH1}
{2\kappa_5^4\alpha^2F_0^2}H_0^2=1 + \frac13\kappa_5^4\alpha^2F_0(R_0F_0-f_0)+\epsilon\sqrt{1+\frac23\kappa_5^4\alpha F_0\big[\alpha(R_0F_0-f_0)\big]}
\end{eqnarray}
where $\epsilon=\pm 1$, the subscript 0 stands for quantities evaluated at the de Sitter space-time, $R_0=12 H_0^2$ and $F=df/dR$. We recover the DGP model for $f(R)=R$. In fact, in that case, the de Sitter self-accelerating DGP branch is obtained for  $\epsilon= 1$ and the normal DGP branch or the  non-self-accelerating solution for $\epsilon=-1$. When the brane action contains curvature corrections to the Hilbert-Einstein action given by the brane scalar curvature, the branch with $\epsilon=-1$ is no longer flat and accelerates (cf. Fig. \ref{ploth}). Therefore, an $f(R)$ term on the brane action induce in a natural way  self-acceleration on the normal branch. Most importantly, it  is known that such a branch is free from the ghost problem (see \cite{Koyama:2007za} and references therein). The reason behind the self-acceleration of the generalised normal brane is the presence of the effective energy density
\begin{equation}
\rho^{(c)}_0=\alpha(F_0R_0-f_0)
\end{equation}
on the modified Friedmann equation on the brane. This can be easily shown by comparing the Friedmann equation (\ref{desitterH1}) with that of modified gravity on brane world-models \cite{BouhmadiLopez:2004ys}

\subsection{Stability of the self-accelerating solutions}

We next analyse the stability of de Sitter solutions under homogeneous perturbations up to first order on $
\delta H= H(t)-H_0$. We will follow the method used in \cite{Faraoni:2005vk}.

The perturbed Friedmann equation (\ref{friedmann}) implies an evolution equation for $\delta H$:
\begin{equation}
\delta \ddot H + 3H_0\delta \dot H + m_{\rm{eff}}^2\delta H=0,
\label{Friedmannpert}
\end{equation}
where $m_{\rm{eff}}^2$ is defined as
\begin{equation}
m_{\rm{eff}}^2=  m_{(4)}^2+m_{\rm{shift}}^2+m_{\rm{pert}}^2.
\label{masseff}
\end{equation}
where
\begin{eqnarray}
 m_{(4)}^2&=&\frac13\left(\frac{F_0}{f_{RR}}-2\frac{f_0}{F_0}\right),\label{othermass}\\
m_{\rm{back}}^2&=&-\frac{2}{\alpha^2\kappa_5^4F_0^2}\left[1-\sqrt{1+\frac23\alpha^2\kappa_5^4F_0(f_0-\kappa_5^2UF_0)}\right],\nonumber \\
m_{\rm{pert}}^2&=&\frac{F_0}{3f_{RR}}\left[1-\sqrt{1+\frac23\alpha^2\kappa_5^4F_0(f_0-\kappa_5^2UF_0)}\right]^{-1}\nonumber.
\end{eqnarray}
and $f_{RR}={d^2f}/{dR^2}$. All this quantities are  evaluated at the de Sitter background solution. Any de Sitter solution is stable as long as $m_{\rm{eff}}^2$ is positive. 

The terms defined on Eq.~(\ref{othermass}) have the following physical meaning: (i) $m_{(4)}^2$ is the analogous quantity to $m_{\rm{eff}}^2$ in a 4D f(R) model \cite{Faraoni:2005vk}, (ii) $m_{\rm{back}}^2$ is a purely background effect due to the shift on the Hubble parameter respect to the standard 4D Case and (iii) $m_{\rm{pert}}^2$ is a purely perturbative extra-dimensional effect.

If we assume that we are close to the 4D regime; i.e. the Hubble rate of the brane is close to its analogous quantity in a 4D $f(R)$ model, then $m_{\rm{back}}^2>0$ and $m_{\rm{pert}}^2<0$. Consequently, $m_{\rm{back}}^2$ tends to make the perturbation heavier. However, the perturbative effect encoded on $m_{\rm{pert}}^2$ would make the perturbation lighter. It can be shown that the extra-dimension has  a \textit{benigner} effect in the 4D f(R) model; i.e. $m_{\rm{eff}}^2>m_{(4)}^2$, as long as\footnote{We have assumed the natural condition $F_0>0$; i.e. the effective gravitational constant of the brane is positive. On the other hand, we have also assumed that we are slightly perturbing the Hilbert-Einstein action of the brane, i.e. $f_0\sim R_0$. Therefore, $f_0$ is positive because $R_0=12H_0^2$.} 
\begin{equation}
{F_0^2}<{4f_0} f_{RR}.
\label{stabilitycondition}
\end{equation}

\section{Conclusions}

We have presented a mechanism to self-accelerate the normal DGP branch which unlike the original self-accelerating DGP branch is known to be free of the ghost problem.  The mechanism is based in including  curvature modifications on the brane action. For simplicity, we choose those terms to correspond to an $f(R)$ contribution, which in addition is known to be the only higher order gravity theories that avoid the so called Ostrogradski instability in 4D models.

\vspace*{0.5cm}\noindent {\bf Acknowledgements}

M.B.L. is  supported by the Portuguese Agency Funda\c{c}\~{a}o para a Ci\^{e}ncia e Tecnologia through the fellowship SFRH/BPD/26542/2006.

\end{document}